\documentclass[a4paper,aps,amsmath,onecolumn,amssymb,nopacs,10pt]{revtex4-2}
\usepackage{epsfig,bm}
\usepackage{graphicx}
\usepackage{amsmath,amssymb,psfrag,textcomp}
\usepackage{amsthm,upgreek}

\usepackage[usenames,dvipsnames,svgnames,table]{xcolor}
\usepackage{hyperref}
\hypersetup{colorlinks=true, linkcolor=NavyBlue, citecolor=PineGreen,urlcolor=cyan}

\newcommand{\bracket}[1]{\left\langle #1\right\rangle}

\newcommand{\be}{\begin{equation}}
\newcommand{\ee}{\end{equation}}
\newcommand{\bd}{\begin{displaymath}}
\newcommand{\ed}{\end{displaymath}}

\newcommand{\beeq}[1] {\begin{equation}\begin{split}#1\end{split}\end{equation}}

\begin{document}
\title{Spectral properties of the generalized diluted Wishart ensemble}
\author{Isaac P\'erez Castillo}
\email{iperez@izt.uam.mx}
\affiliation{Departamento de F\'isica, Universidad Autónoma Metropolitana-Iztapalapa, San Rafael Atlixco 186, Ciudad de México 09340, Mexico}
\begin{abstract}
The celebrated Mar\v{c}enko-Pastur law, that considers the asymptotic spectral density of random covariance matrices, has found a great number of applications in physics, biology, economics, engineering, among others. Here, using techniques from statistical mechanics of spin glasses, we derive simple formulas concerning the spectral density of generalized diluted  Wishart matrices. These are defined as $\bm{F}\equiv \frac{1}{2d}\left( \bm{X}\bm{Y}^T+ \bm{Y}\bm{X}^T\right)$, where $\bm{X}$ and $\bm{Y}$ are diluted  $N\times P$ rectangular matrices, whose entries correspond to the links of  doubly-weighted random bipartite Poissonian graphs following the distribution $P(x_{i}^\mu,y_{i}^{\mu})=\frac{d}{N}\varrho(x_{i}^{\mu},y_{i}^{\mu})+\left(1-\frac{d}{N}\right)\delta_{x_{i}^{\mu},0}\delta_{y_{i}^{\mu},0}$, with the probability density $\varrho(x,y)$ controlling the correlation between the matrices entries of $\bm{X}$ and $\bm{Y}$. Our results cover several interesting cases by varying the parameters of the matrix ensemble, namely, the dilution of the graph $d$, the rectangularity of the matrices $\alpha=N/P$,  and the degree of correlation of the matrix entries via the density $\varrho(x,y)$. Finally, we compare our findings to numerical diagonalisation  showing excellent agreement.
\end{abstract}
\maketitle

\section{Introduction}
The Wishart distribution was introduced in 1928 by John Wishart  \cite{Wishart1928} and it can be thought of as a generalization to a higher dimensional gamma distribution. Along the years it has become one of the central matrix ensembles in random matrix theory, the so-called Wishart ensemble. It naturally arises as the probability function of sample covariance matrices for a sample of multivariate normal distributions and, as such, has found a great number of applications in the context of multivariate statistical analysis.

From the several statistical properties derived from the Wishart ensemble, the limiting distribution of its spectral density, the celebrated Mar\v{c}enko-Pastur \cite{Marcenko1967} law, can be ubiquitously found in multiple applications ranging from physics, to biology, to economics, to engineering, among others. More precisely, let us consider a set  $\{x_i^{\mu}\}$ of independent and identically distributed random variables with zero mean and unit variance, with $i=1,\ldots,N$ and $\mu=1,\ldots, P$.  This may be understood as $N$ times series each of size $P$. The entries of the $N\times N$ sample covariance matrix $\bm{E}$ are defined as:
\beeq{
(\bm{E})_{ij}\equiv \frac{1}{P}\sum_{\mu=1}^P x^{\mu}_i x^{\mu}_j=\frac{1}{P}(\bm{X}\bm{X}^T)_{ij}\,,
}
where we have introduced an $N\times P$ matrix $\bm{X}$ whose elements are precisely $(\bm{X})
_{i\mu}=x^\mu_i$. Notice that in the limit $P\to \infty$, while keeping $N$ fixed, the covariance matrix becomes the identity matrix and its spectral density is a Dirac delta distribution centered at one. However, in the limiting case of  $P\to \infty$ and  $N\to \infty$, while keeping the ratio $\alpha=N/P$ fixed, the spectral density $\rho(\lambda)$ is described the Mar\v{c}enko-Pastur law \cite{Marcenko1967}:
\beeq{
\rho(\lambda)=\frac{1}{2\pi \alpha \lambda}\sqrt{(\lambda_{+}-\lambda)(\lambda-\lambda_{-})}\openone_{\lambda\in[\lambda_{-},\lambda_{+}]}\,,
}
with $\lambda_{\pm}=\left(1\pm \sqrt{\alpha}\right)^2$.

This ensemble was recently generalized in \cite{Bouchaud2020} by considering the  symmetric cross-correlation ensemble, composed of matrices of the following form:
\beeq{
\bm{F}\equiv \frac{1}{2P}\left(\bm{X}\bm{Y}^T+\bm{Y}\bm{X}^T\right)\,,
}
where $\bm{X}$ and $\bm{Y}$ are two $N\times P$ rectangular matrices, each with independent and identically distributed zero mean and unit variance entries, but such that $\mathbb{E}[x_i^\mu y_j^\mu]=c\delta_{ij}$, so that $c\in [-1,1]$ controls the correlation between the elements of both matrices. Obviously, for $c=1$ one recovers the classical Wishart ensemble. The ensemble, as pointed out in \cite{Bouchaud2020}, can be seen, for instance, as a null hypothesis for detecting correlations between different multidimensional time series, or as the statistical properties of the anticommutator of random matrices for $N=P$. Its spectral density was derived in \cite{Bouchaud2020} using free probability theory.

On the other hand, random graphs constitute the main tool to model the complex behavior of large empirical networks observed in social, technological, and biological systems \cite{Newman2018,Dorogovtsev2003}. In random graph models a network is typically represented by nodes that interact through edges. Random graph theory leads to important insights into the structure of networks as well as on the dynamical processes occurring on them, such as the spreading of diseases \cite{Van_Mieghem2012,Goltsev2012}, the stability of ecosystems to perturbations \cite{Neri2020},  the dynamics and equilibrium properties of sparsely connected neurons \cite{Brunel2000,Castillo2004,Castillo2004b}, models  of formation of opinion  \cite{Giuraniuc2005}, and so on. Being the spectral properties of random graphs a central property to understand these systems, in a  series of seminal works the spectral density of standard, fully connected, random matrix ensembles was generalised to diluted random graphs, first for symmetric matrices in \cite{Rogers2008}, subsequently for asymmetric ones \cite{Rogers2009}, which was later followed by correlated networks \cite{Rogers2010}. Moreover, recently a set of powerful mathematical methods have also been introduced to obtain the large deviation properties of diluted random matrix ensembles \cite{Metz2016,Metz2017,Castillo2018}.

The main goal of the present work is to introduce the diluted symmetric cross-correlation ensemble and to study its spectral density using mathematical tools originated in statistical mechanics of spin glasses \cite{Mezard1987}. This work is organized as follows. In Sect. \ref{sect:definitions} we start by defining this ensemble, discuss how by changing the parameters of the ensemble we cover previous cases, and go through the main derivations to obtain the spectral density by using the cavity method. In Sect. \ref{sect:results} we summarize our theoretical results, exploring some particular cases,  and compare them to numerical diagonalization. We conclude with a summary and discuss future lines of research.

\section{Definitions and main theoretical derivations}
\label{sect:definitions}
We start by introducing the diluted symmetric cross-correlation ensemble, which we will also call as the generalized diluted Wishart ensemble. To do this, let us take two $N\times P$ rectangular matrices $\bm{X}$ and $\bm{Y}$ and use them to introduce an $N\times N$ matrix $\bm{F}$ given by
\beeq{
\bm{F}=\frac{1}{2d}\left(\bm{X}\bm{Y}^T+ \bm{Y}\bm{X}^T\right)\,.
}
Next, we assume that the matrix entries of $\bm{X}$ and $\bm{Y}$  weigh the double links  of a Poisson bipartite graph with probability distribution:
\beeq{
P(x_{i}^{\mu},y_{i}^{\mu})=\frac{d}{N}\varrho(x_{i}^{\mu},y_{i}^{\mu})+\left(1-\frac{d}{N}\right)\delta_{y_{i}^{\mu},0}\delta_{y_{i}^{\mu},0}\,,
}
where the distribution function $\varrho(x_{i}^{\mu},y_{i}^{\mu})$ controls the correlation between the matrix entries $x_{i}^{\mu}$ and $y_{i}^{\mu}$ between a pair of connected nodes $i$ and $\mu$.  Note that if we take $d=N$ we recover the fully connected or dense limit studied in \cite{Bouchaud2020}. If the distribution function $\rho$ is such that  the variables $x_{i}^{\mu}$ and $y_i^{\mu}$ are perfectly correlated, one recovers the original Wishart ensemble.

\begin{figure}[h]
\includegraphics[width=14cm,height=10cm]{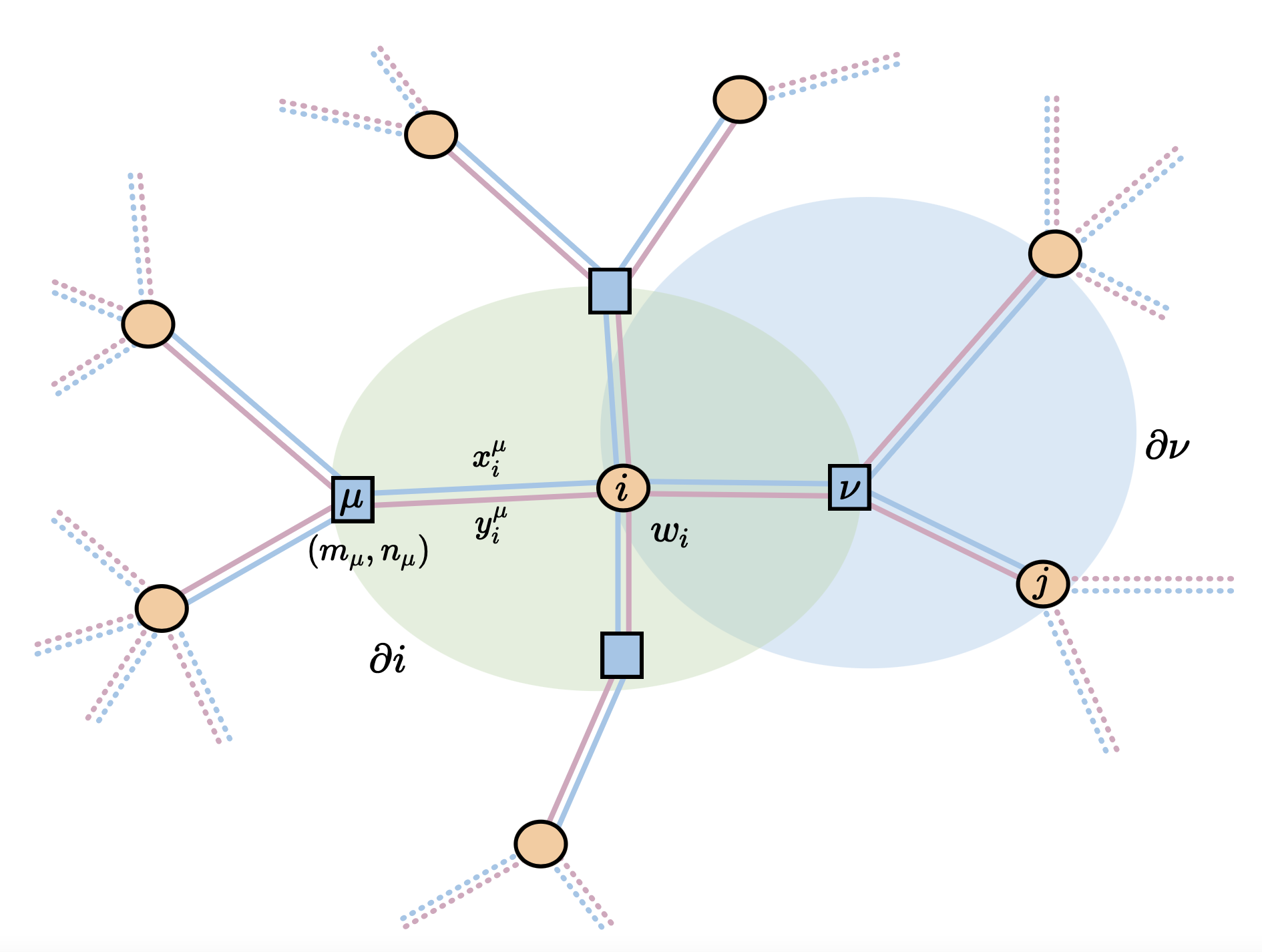}
\caption{Bipartite graph representation of a locally-tree like matrix $\bm{F}$}
\label{fig1}
\end{figure}

Let $\{\lambda^{(\bm{F})}_{i}\}_{i=1}^N$ be the spectrum of a matrix $\bm{F}$ and  recall that its empirical spectral density is given by
\beeq{
\rho_{\bm{F}}(\lambda)=\frac{1}{N}\sum_{i=1}^N\delta(\lambda-\lambda^{(\bm{F})}_i)\,.
}
Then, following \cite{Edwards1976,Rogers2008}, we recast this as a spin-glass-type problem in which the $\rho_{\bm{F}}(\lambda)$ is given by
\beeq{
\rho_{\bm{F}}(\lambda)=-\lim_{\epsilon\to 0^+}\frac{2}{\pi N}\text{Im}\left[\frac{\partial}{\partial z}\log Z_{\bm{F}}(z)\right]_{z=\lambda -i\epsilon}
}
where $Z_{\bm{F}}(z)$ can be thought of as a partition function of a spin glass model:
\beeq{
Z_{\bm{F}}(z)=\int\left[\prod_{i=1}^N\frac{dw_i}{\sqrt{2\pi}}\right] e^{-H_{\bm{F}}(\bm{w};z)}\,,
}
where we have introduced the following Hamiltonian:
\beeq{
H_{\bm{F}}(\bm{w};z)=\frac{z}{2}\sum_{i=1}^N x_i^2-\frac{1}{2}\sum_{\mu=1}^P n_\mu(x_{\partial \mu})m_\mu(x_{\partial \mu})\,,
}
with
\beeq{
n_\mu(x_{\partial \mu})=\frac{1}{\sqrt{d}}\sum_{i\in\partial \mu}x_{i}^{\mu} w_i\,,\quad\quad m_\mu(x_{\partial \mu})=\frac{1}{\sqrt{d}}\sum_{i\in\partial \mu}y_{i}^{\mu} w_i\,.
}
In this context, the empirical spectral density is given by the statistical average of the $w$-variables via the formula:
\beeq{
\rho_{\bm{F}}(\lambda)=\lim_{\epsilon\to 0^{+}}\text{Im}\left[\frac{1}{\pi N}\sum_{i=1}^N\bracket{w^2_i}_{{z=\lambda -i\epsilon}}\right]
\label{eq:1}\,,
}
with
\beeq{
\bracket{(\cdots)}_z=\frac{1}{Z_{\bm{F}}(z)}\int\left[\prod_{i=1}^N\frac{dw_i}{\sqrt{2\pi}}\right] e^{-H_{\bm{F}}(\bm{w};z)}(\cdots)\,.
}
As, according to Eq. $\eqref{eq:1}$, the empirical spectral density depends on the expectation value on single nodes, it seems natural to apply the cavity method to find a closed set of equations for single-node marginals. To do so, we notice that the Hamiltonian $H_{\bm{F}}(\bm{w};z)$ models interacting variables on an underlying bipartite graph as shown in Fig. \ref{fig1}.  The graph has two types of nodes: factor nodes or $\mu$-nodes, represented by squares and labelled by greek letters; graph nodes or $i$-nodes, represented by circles and labelled by latin letters. Furthermore, on the $i$-nodes we have dynamical variables, denoted as $\bm{w}=(w_1,\ldots,w_N)$, while on the factor nodes we have a pair of dynamical variables $\{\bm{m},\bm{n}\}=\{(m_1,\ldots,m_P),(n_1,\ldots,n_P)\}$. The graphs nodes are connected to the factor nodes, and vice versa, by  pairs of links which correspond to the matrix entries of $\bm{X}$ and $\bm{Y}$. As usual the neighbourhood of a graph node, let us say $i$, is denoted as $\partial i$, and similarly, the neighbourhood of a factor node, say $\mu$, is denoted as $\partial \mu$. Finally, given a subset $A$ of nodes, $w_A\equiv \{w_\ell\}_{\ell\in A}$ and $d w_A\equiv\prod_{\ell\in A} dw_\ell$.

After performing the conventional derivation using the cavity method (see, for instance, \cite{Rogers2008}), we find that single-node marginals $P_i(w_i)$ are given by:
\beeq{
P_i(w_i)=\frac{e^{-\frac{z}{2}x_i^2}}{Z_i}\int dm_{\partial i\setminus \mu}dn_{\partial i\setminus \mu} e^{\frac{1}{2}\sum_{\nu\in\partial i} \left(m_\nu+\frac{1}{\sqrt{d}} x_{i}^{\nu} w_i\right)\left(n_\nu+\frac{1}{\sqrt{d}}y_{i}^{\nu} w_i\right)}\prod_{\nu\in \partial i} Q_\nu^{(i)}(m_\nu, n_\nu)\,,\quad i=1,\ldots,N\,,
\label{eq2}
}
where the so-called  cavity marginals $Q_\nu^{(i)}(m_\nu, n_\nu)$ obey the following closed set of  equations
\beeq{
Q_\nu^{(i)}(m_\nu, n_\nu)&=\frac{1}{Z_\nu^{(i)}}\int dw_{\partial \nu\setminus i}\delta\left(m_\nu -\frac{1}{\sqrt{d}}\sum_{\ell\in \partial \nu\setminus i}x_{\ell}^{\nu}w_\ell\right)\\
&\times\delta\left(n_\nu -\frac{1}{\sqrt{d}}\sum_{\ell\in \partial \nu\setminus i} y_{\ell}^{\nu}w_\ell\right)\prod_{\ell\in \partial \nu\setminus i}P_\ell^{(\nu)}(w_\ell)\,,\quad \nu=1,\ldots,P\,,\quad  i\in\partial \nu\,,\\
P_i^{(\mu)}(w_i)&=\frac{e^{-\frac{z}{2}x_i^2}}{Z_i^{(\mu)}}\int dm_{\partial i\setminus \mu}dn_{\partial i\setminus \mu} e^{\frac{1}{2}\sum_{\nu\in\partial i\setminus \nu} \left(m_\nu+\frac{1}{\sqrt{d}} x_{i}^{\nu} w_i \right)\left(n_\nu+\frac{1}{\sqrt{d}} y_{i}^{\nu} w_i\right)}\\
&\times\prod_{\nu\in \partial i\setminus \mu} Q_\nu^{(i)}(m_\nu, n_\nu)\,,\quad i=1,\ldots N\,,\quad \mu\in\partial i\,.
\label{eq3}
}
As it was highlighted in \cite{Rogers2008}, the closed set of  equations given by \eqref{eq3} can be further simplified by noticing that the set of Gaussian distributions is a fixed point. With this in mind, and with a modest amount of foresight,  we write
\beeq{
P_\ell ^{(\nu)}(w_\ell)&=\frac{1}{\sqrt{2\pi \Delta_\ell^{(\nu)}}} e^{-\frac{w_\ell^2}{2\Delta_\ell^{(\nu)}}}\,,\\
Q_\nu^{(i)}(m_\nu,n_\nu)&=\frac{1}{\sqrt{(2\pi)^2\det G_\nu^{(i)}}}\exp\left[-\frac{1}{2} (m_\nu, n_\nu) [G_{\nu}^{(i)}]^{-1}\begin{pmatrix}m_\nu\\n_\nu\end{pmatrix}\right]\,,\quad
G_\nu^{(i)}=\begin{pmatrix}
g_\nu^{(i)}&\gamma_\nu^{(i)}\\
\gamma_\nu^{(i)}& \delta_\nu^{(i)}
\end{pmatrix}\,.
\label{eq3b}
}
After plugging Eqs. \eqref{eq3b} into the set of Eqs. \eqref{eq3}, and a bit of algebra, we arrive to the following set of equations
\beeq{
 \Delta_{i}^{(\mu)}&=\frac{1}{z-\frac{1}{d}\sum_{\nu\in \partial i\setminus \mu} \frac{ [x_{i}^{\nu}]^2 g_\nu^{(i)}+ [y_{i}^{\nu}]^2\delta_\nu^{(i)}+2x_{i}^{\nu}y_{i}^{\nu}(2-\gamma_\nu^{(i)})}{(\gamma_\nu^{(i)}-2)^2-\delta_\nu^{(i)} g_\nu^{(i)}}}\,,\\
g_\nu^{(i)}&=\frac{1}{d}\sum_{\ell\in \partial \nu\setminus i}[x_{\ell}^{\nu}]^2\Delta_\ell^{(\nu)}\,,\quad\gamma_\nu^{(i)}=\frac{1}{d}\sum_{\ell\in \partial \nu\setminus i}x_{\ell}^{\nu} y_{\ell}^{\nu}\Delta_\ell^{(\nu)}\,,\quad 
\delta_\nu^{(i)}=\frac{1}{d}\sum_{\ell\in \partial \nu\setminus i}[y_{\ell}^{\nu}]^2\Delta_\ell^{(\nu)}\,.
\label{eq:final1}
}
 Once we have found a solution to the set of Eqs. \eqref{eq:final1}, numerically or otherwise, the spectral density is given by:
\beeq{
\rho(\lambda)=\lim_{\epsilon\to0^{+}}\frac{1}{\pi N}\sum_{i=1}^N \Delta_i(\lambda-i\epsilon)\,,
\label{eq:final2}
}
with
\beeq{
 \Delta_{i}(z)&=\frac{1}{z-\frac{1}{d}\sum_{\nu\in \partial i} \frac{ [x_{i}^{\nu}]^2 g_\nu^{(i)}+ [y_{i}^{\nu}]^2\delta_\nu^{(i)}+2x_{i}^{\nu}y_{i}^{\nu}(2-\gamma_\nu^{(i)})}{(\gamma_\nu^{(i)}-2)^2-\delta_\nu^{(i)} g_\nu^{(i)}}}\,.
\label{eq:final3} 
}
The set of equations \eqref{eq:final1}, \eqref{eq:final2}, and \eqref{eq:final3} are the first main result of this paper and correspond the empirical spectral density based on the cavity equations. Note that when applying the cavity method to derive these equations one assumes that the underlying graph is a typical large locally tree-like graph, being Poissonian graphs one of several possibilities.  For Poissonian graphs the average ensemble of the cavity equations acquires a simpler form. Indeed, following \cite{Mezard2001} to carry out the ensemble average of the cavity equations, we arrive to the following set of self-consistency equations:
\beeq{
w(g,\gamma,\delta)&=\sum_{\ell=0}^\infty\frac{e^{-d} d^\ell }{\ell!}\int\left[\prod_{k=1}^\ell v(\Delta_k) d \Delta_k \varrho(x_k,y_k) dx_k dy_k\right]\\
&\times\delta\left(g-\frac{1}{d}\sum_{k=1}^\ell x^2_{k}\Delta_k\right)
\delta\left(\gamma-\frac{1}{d}\sum_{k=1}^\ell x_{k} y_{k}\Delta_k\right)
\delta\left(\delta-\frac{1}{d}\sum_{k=1}^\ell y^2_{k}\Delta_k\right)\\
v(\Delta)&=\sum_{\ell=0}^\infty\frac{e^{-\alpha d} (\alpha d)^\ell }{\ell!}\int\left[\prod_{k=1}^\ell w(g_k,\gamma_k,\delta_k)   dg_k d\gamma_k d\delta_k\varrho(x_k,y_k) dx_k dy_k\right]\\
&\times\delta\left(\Delta-\frac{1}{z-\frac{1}{d}\sum_{k=1}^\ell \frac{ x^2_{k} g_k+ y_{k}^2\delta_k+2x_{k}y_{k}(2-\gamma_k)}{(\gamma_k-2)^2-\delta_k g_k}}\right)
\label{eq:popeqs}
}
where $w(g,\gamma,\delta)$ and $v(\Delta)$ are densities. In this case the spectral density is given by instead by the following expression:
\beeq{
\rho(\lambda)=\lim_{\epsilon\to 0^{+}}\frac{1}{\pi} \int d \Delta v(\Delta) \text{Im}[\Delta(z)]_{z=\lambda-i\epsilon}\,.
}
\section{Analysis of the theoretical results and comparison with numerical estimates}
\label{sect:results}
Let us start first by considering the dense limit consisting on taking the limit $d\to \infty$ while keeping $\alpha$ fixed. There are two ways to do this, both instructive and complementary: either using the set of cavity equations \eqref{eq:final1} or the ensemble average equations \eqref{eq:popeqs}. We take the first route. Consider for instance the relationship between the coefficients $g_\nu^{(i)}$ and the $\Delta_{\ell}^{(\nu)}$. The Onsager's correction term \cite{Mezard1987} between $g_\nu^{(i)}$ and $g_\nu$, and similarly for the $\Delta$ variables, is order $\mathcal{O}(d^{-1})$ and vanishes in the dense limit (see, for instance, \cite{Rogers2008}). Moreover the coefficients $\Delta_{\ell}^{(\nu)}$ cannot depend on the $\{x_\ell^\nu\}$ random variables since for that coefficient the factor node $\nu$ has been removed. Thus we can graciously write that
\beeq{
\lim_{d\to\infty} \frac{1}{d}\sum_{\ell\in \partial \nu\setminus i}[x_{\ell}^{\nu}]^2\Delta_\ell^{(\nu)}=\mathbb{E}[(x^\nu_\ell)^2]\lim_{d\to\infty}\frac{1}{d}\sum_{\ell=1}^d \Delta_\ell\equiv \Delta\,.
}
We similarly arrive at the following limits
\beeq{
\lim_{d\to\infty}\frac{1}{d}\sum_{\ell\in \partial \nu\setminus i}x_{\ell}^{\nu} y_{\ell}^{\nu}\Delta_\ell^{(\nu)}&=\mathbb{E}[x_{\ell}^{\nu} y_{\ell}^{\nu}]\lim_{d\to\infty}\frac{1}{d}\sum_{\ell=1}^d\Delta_\ell=c\Delta\,,\\ 
\lim_{d\to\infty}\frac{1}{d}\sum_{\ell\in \partial \nu\setminus i}[y_{\ell}^{\nu}]^2\Delta_\ell^{(\nu)}&=\mathbb{E}[(y_{\ell}^{\nu})^2]\lim_{d\to\infty}\frac{1}{d}\sum_{\ell=1}^d\Delta_\ell=\Delta\,.
}
This yields the following equations for $\Delta$:
\beeq{
\Delta&=\frac{1}{z-\alpha \frac{ 2\Delta+2c(2-c\Delta)}{(c\Delta -2)^2-\Delta^2}}\,,
\label{eq:denselimit}
}
from where the spectral density is obtained by $\rho(\lambda)=\lim_{\epsilon\to 0^{+}}\text{Im}\Delta(\lambda-i\epsilon)/\pi$, in agreement with the results found in \cite{Bouchaud2020}. Notice that in the limit $c\to 1$ this equation reduces to
\beeq{
\Delta&=\frac{1}{z- \frac{\alpha}{1-\Delta}}\,,
}
which is the propagator equation for the Mar\v{c}enko-Pastur spectral density. For the case $c=0$ we have instead 
\beeq{
\Delta=\frac{1}{z-\alpha\frac{2\Delta}{4-\Delta^2}}\,.
}
These cases were thorouhgly analised in \cite{Bouchaud2020} and therefore we will not dwell around them any longer.

To obtain the spectral density for the general case of finite $d$, the set of Eqs. \eqref{eq:popeqs} is solved by using population dynamics, a method that combines Monte Carlo integration and fixed point iteration method. To apply it, each density $w(g,\gamma,\delta)$ and $v(\Delta)$ is represented by a population of $\mathcal{N}$ random variables $\{g_a,\gamma_a,\delta_a\}_{a=1}^{\mathcal{N}}$ and $\{\Delta_a\}_{a=1}^{\mathcal{N}}$, respectively, whose histograms are precisely  estimates of the aforementioned densities and they become more and more accurate as the population size $\mathcal{N}$ increases. More precisely,
\beeq{
v(\Delta)&=\lim_{\mathcal{N}\to\infty}\frac{1}{\mathcal{N}}\sum_{\alpha=1}^{\mathcal{N}}\delta(\Delta-\Delta_\alpha)\,,\\
w(g,\gamma,\delta)&=\lim_{\mathcal{N}\to\infty}\frac{1}{\mathcal{N}}\sum_{\alpha=1}^{\mathcal{N}}\delta(g-g_\alpha)\delta(\gamma-\gamma_\alpha)\delta(\delta-\delta_\alpha)\,.
}
Then, starting with a random population  $\{g_a,\gamma_a,\delta_a\}_{a=1}^{\mathcal{N}}$ and $\{\Delta_a\}_{a=1}^{\mathcal{N}}$, these are updated according to the following steps:
\begin{enumerate}
\item Generate a Poissson random number $\ell$ with mean value $d$ and select randomly and uniformly $\ell$ elements from the population $\{\Delta_a\}_{a=1}^{\mathcal{N}}$.
\item Select uniformly and randomly three elements $\{g,\gamma,\delta\}$ from the population $\{g_a,\gamma_a,\delta_a\}_{a=1}^{\mathcal{N}}$ and replace them with the new values $\frac{1}{d}\sum_{\ell\in \partial \nu\setminus i}[x_{\ell}^{\nu}]^2 \Delta_\ell^{(\nu)}$, $\frac{1}{d}\sum_{\ell\in \partial \nu\setminus i}x_{\ell}^{\nu} y_{\ell}^{\nu}\Delta_\ell^{(\nu)}$, and $\frac{1}{d}\sum_{\ell\in \partial \nu\setminus i}[y_{\ell}^{\nu}]^2\Delta_\ell^{(\nu)}$, respectively.
\item Generate a Poisson random number $\ell$ with mean value $\alpha d$ and select randomly and uniformly $\ell$ elements from the population $\{g_a,\gamma_a,\delta_a\}_{a=1}^{\mathcal{N}}$.
\item Select uniformly and randomly one element of the population $\{\Delta_a\}_{a=1}^{\mathcal{N}}$ and replace it with the new value $[z-\frac{1}{d}\sum_{k=1}^\ell \frac{ x^2_{k} g_k+ y_{k}^2\delta_k+2x_{k}y_{k}(2-\gamma_k)}{(\gamma_k-2)^2-\delta_k g_k}]^{-1}$.
\item Iterate until convergence. 
\end{enumerate} 
Convergence is usually monitored by using any stopping criterion commonly found in numerical iteration methods.
\begin{figure}[t]
\begin{center}
\includegraphics[height=5cm, width=5cm]{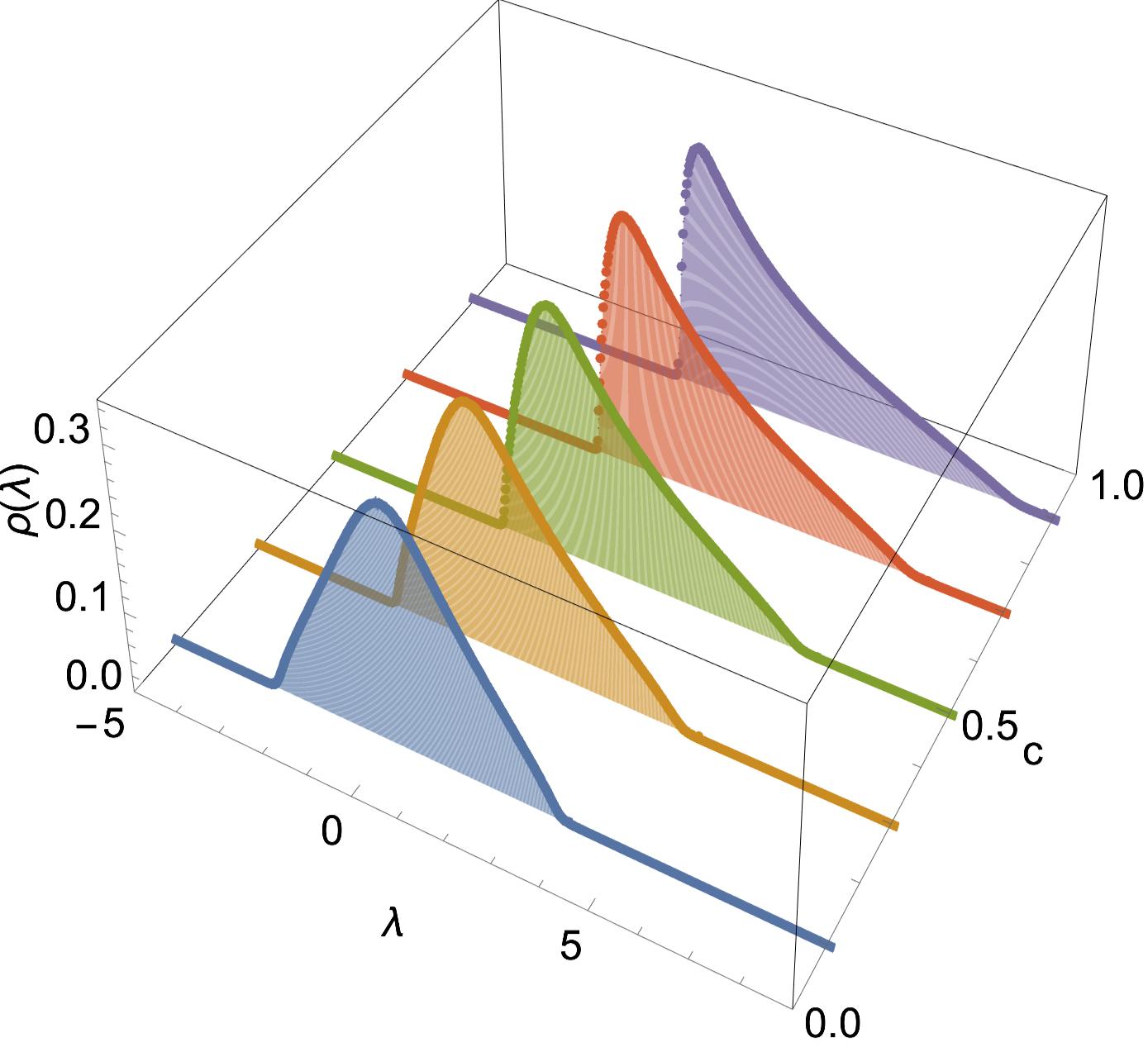}\quad
\includegraphics[height=5cm, width=5cm]{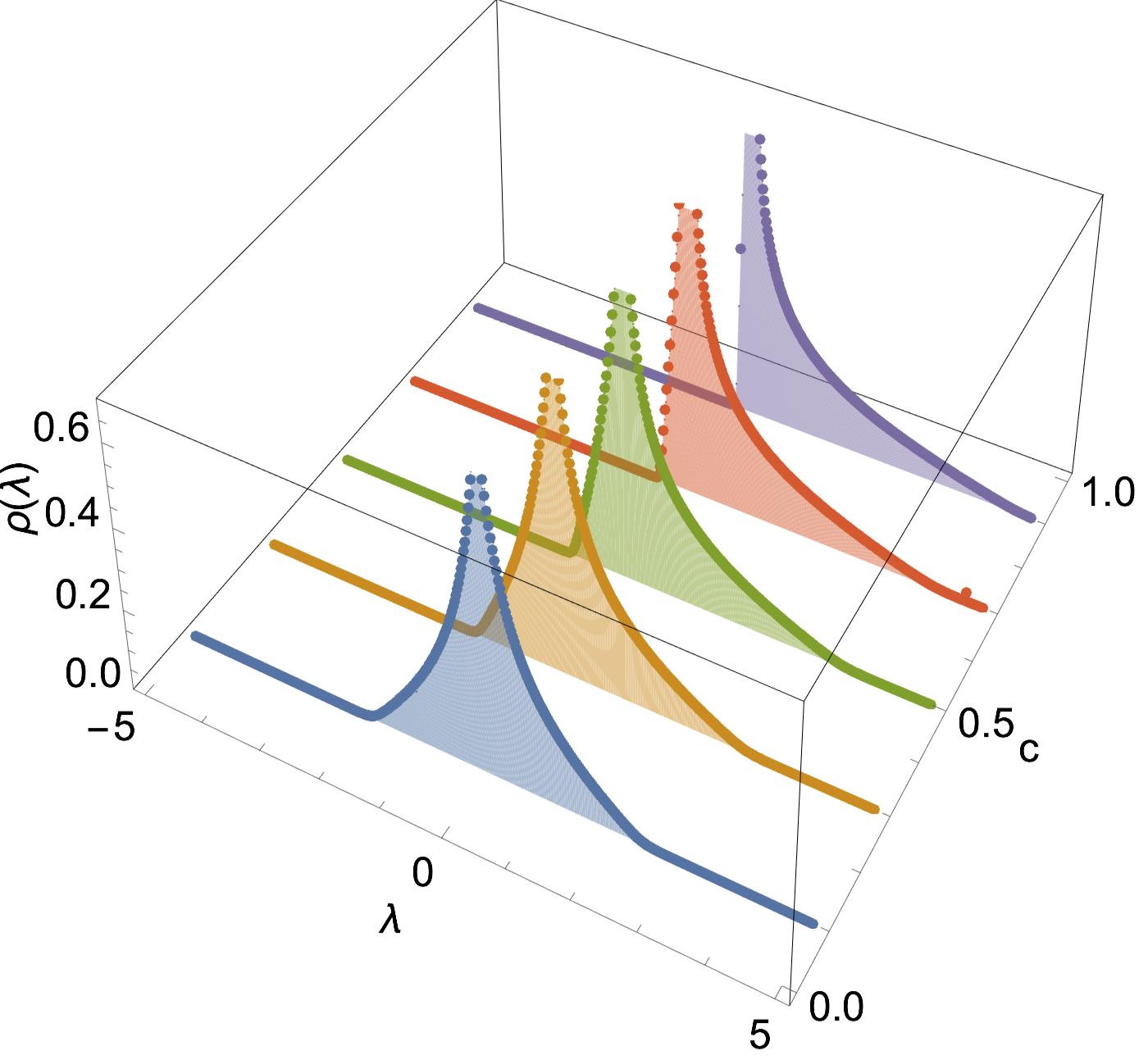}\quad
\includegraphics[height=5cm, width=5cm]{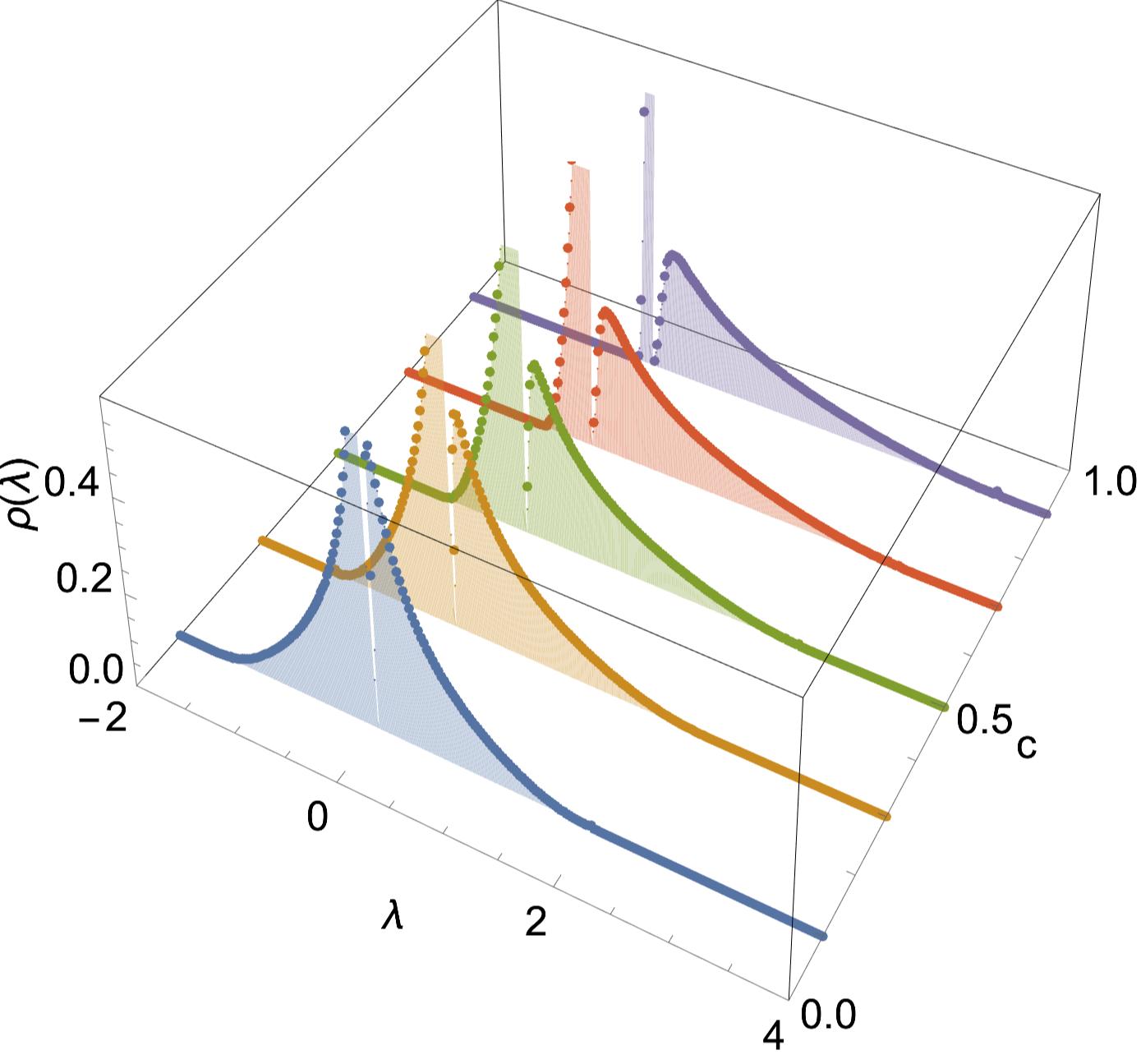}
\end{center}
\caption{Comparison between the theorerical results (markers) and numerical diagonalization (histograms filling the plots) for $d=12$ a values of $c$ from 0.1 until 0.9. The parameter $\alpha$, that controls the rectangularity of the matrix, is fixed to the values $\alpha=3$ (left figure), $\alpha=1$ (middle figure), and $\alpha=0.3$ (right figure).}
\label{fig2}
\end{figure}

In Fig. \ref{fig2}, we compare the theoretical results using population dynamics (represented with  markers) with numerical diagonizalisation (represented by the histograms that fill each curve to the bottom). For the former we use a population size of $\mathcal{N}=10^5$ while for the numerics we diagonizalise  a set of  $10^3$ matrices of size $N=10^3$. For the comparison we take  $d=12$ and  $\alpha\in\{3, 1, 0.3\}$ (from left to right figures, respectively) and, in each plot, for varying values of the correlation $c=\mathbb{E}[x_i^\mu y_i^\mu]$. As we can see, the agreement between theory and numerics is excellent. Generally speaking the spectral density does not have hard edges and the end points of its domain. When the rectangularity parameter $\alpha>1$, the spectral density $\rho(\lambda)$ is bounded, while for $\alpha\leq 1$ it develops a gap and a singularity at $\lambda=0$.

\section{Conclusions}
\label{sect:conclusions}
Being random graphs one of the main tools to model complex phenomena, in this paper we have introduced the diluted version of the symmetric cross-correlation matrix ensemble and have obtained its limiting spectral density exactly by using the cavity method. We have checked that our results are correct by thoroughly comparing them to estimates of the empirical density obtained by numerical diagonalization. \\

This work opens the door to study other spectral properties of this ensemble as, for instance, the distribution of extreme eigenvalues,  rate functions using large deviation theory, and to explore asymmetric versions of this matrix ensemble. These and some other studies are currently under way.

\bibliographystyle{apsrev4-2}
\bibliography{biblio.bib}

\end{document}